\documentclass[final,5p,times,twocolumn]{elsarticle}

\usepackage{graphicx}
\usepackage{epsfig}
\usepackage{amssymb}

\newcommand{\lnsco}{La$_{1.6-x}$Nd$_{0.4}$Sr$_x$CuO$_4$}

\journal{Physica C}

\begin{document}

\begin{frontmatter}

\title{Zooming on the Quantum Critical Point in Nd-LSCO}

\author[UdeS]{Olivier Cyr-Choini\`{e}re}
\author[UdeS,Dresden]{R. Daou}
\author[UdeS]{J. Chang}
\author[UdeS]{Francis Lalibert\'{e}}
\author[UdeS]{Nicolas Doiron-Leyraud} 
\author[UdeS]{David LeBoeuf}

\author[NHMFL]{Y. J. Jo}
\author[NHMFL]{L. Balicas}

\author[Ames]{J.-Q. Yan}
\author[UTex]{J.-G. Cheng}
\author[UTex]{J.-S. Zhou}
\author[UTex]{J. B. Goodenough}

\author[UdeS,CIFAR,email]{Louis Taillefer}

\address[UdeS]{D\'{e}partement de physique and RQMP, Universit\'{e} de
  Sherbrooke, Sherbrooke,  Qu\'{e}bec J1K 2R1, Canada.}
\address[Dresden]{Present address: Dresden High Magnetic Field Laboratory, 01328 Dresden, Germany.}
\address[NHMFL]{National High Magnetic Field Laboratory, Florida State University, 
	Tallahassee, Florida 32310-3706, USA.}
\address[Ames]{Ames Laboratory, Ames, Iowa 50011, USA}
\address[UTex]{Texas Materials Institute, University of Texas at Austin, 
  Austin, Texas 78712, USA.}
\address[CIFAR]{Canadian Institute for Advanced Research, Toronto, Ontario M5G 1Z8, Canada.}
\address[email]{E-mail: louis.taillefer@physique.usherbrooke.ca}

\begin{abstract}

Recent studies of the high-$T_{\rm c}$ superconductor \lnsco{} (Nd-LSCO) have found a linear-$T$ 
in-plane resistivity $\rho_{\rm ab}$ and a logarithmic temperature dependence of the thermopower 
$S\,/\,T$ at a hole doping $p = 0.24$, and a Fermi-surface reconstruction just below $p = 0.24$~\cite{daou08,daou09}. 
These are typical signatures of a quantum critical point (QCP). Here we report data on the 
$c$-axis resistivity $\rho_{\rm c}(T)$ of Nd-LSCO measured as a function of temperature near 
this QCP, in a magnetic field large enough to entirely suppress superconductivity. Like 
$\rho_{\rm ab}$, $\rho_{\rm c}$ shows an upturn at low temperature, a signature of Fermi 
surface reconstruction caused by stripe order. Tracking the height of the upturn as it 
decreases with doping enables us to pin down the precise location of the QCP where stripe 
order ends, at $p^\star = 0.235 \pm 0.005$.We propose that the temperature $T_{\rm \rho}$ below 
which the upturn begins marks the onset of the pseudogap phase, found to be roughly twice 
as high as the stripe ordering temperature in this material.

\end{abstract}


\end{frontmatter}

One of the central questions of high-$T_{\rm c}$ superconductivity is the nature of the 
pseudogap phase. Recent quantum oscillation studies~\cite{doiron07} favour a scenario of 
competing order, as they reveal that the large hole-like Fermi surface of overdoped 
cuprates~\cite{vignolle08} transforms into small electron-like pockets in the pseudogap phase~\cite{leboeuf07}. 
This shows that there is some ``hidden'' order in the pseudogap phase which breaks 
translational symmetry and thus causes a reconstruction of the Fermi surface ~\cite{chakravarty08}. 
In some cuprates, such as ~\lnsco{} (Nd-LSCO), there is clear evidence for charge / spin order, 
better known as ``stripe order'', setting in at low temperature (see Ref.~\cite{hunt01} 
and references therein), and the pseudogap phase may be a precursor to that stripe phase.

The presence of an order in the phase diagram involves the presence of a 
quantum critical point (QCP) at a critical doping $p^\star$ where the ordering 
temperature goes to zero. In Nd-LSCO at a hole-doping of $p = 0.24$, 
the in-plane resistivity $\rho_{\rm ab}$ is linear down to the lowest temperature~\cite{daou08}, 
and the thermopower $S$ has a $T\ln(1/T)$ dependence over a decade of temperature~\cite{daou09}. 
These are typical signatures of a quantum phase transition for a metal with 
two-dimensional antiferromagnetic fluctuations~\cite{lohneysen07,paul01}.
They show that the QCP is close to but slightly below that doping, 
{\it i.e.} $p^\star < 0.24$.

\begin{figure}
	\begin{center}
		\epsfig{figure=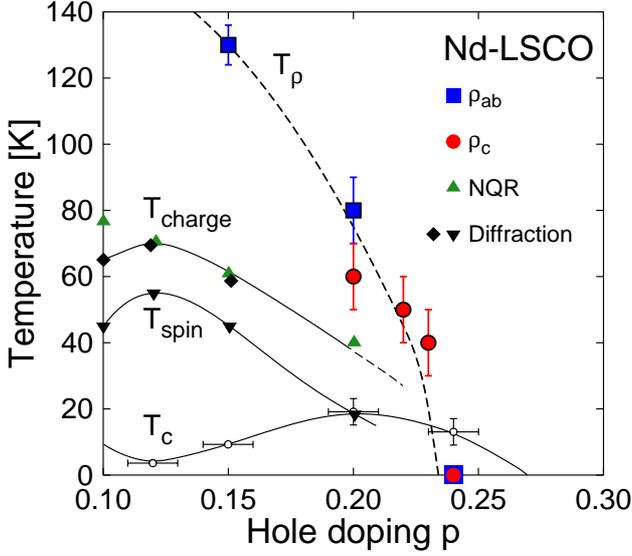, angle=270,width=0.46\textwidth}
			\caption{(Color online) Temperature-doping phase diagram of Nd-LSCO. 
			  The temperature $T_{\rm \rho}$ is defined 
  			as the temperature below which the normal-state resistivity $\rho(T)$ 
  			deviates from its linear-$T$ behavior (blue squares for in-plane, 
  			red circles for $c$-axis). Data for $p = 0.15$ is 
  			from~\cite{ichikawa00}; $\rho_{\rm ab}$ data for $p = 0.20$ and $p = 0.24$ are from~\cite{daou08}.
  			At $p=0.24$, $T_\rho = 0$ because there is no deviation from linearity down to the lowest 
  			temperature \cite{daou08}. 
  			We also show $T_{\rm charge}$, the onset temperature for charge order, detected either via 
  			nuclear quadrupole resonance (NQR; green up triangles) or
  			via X-ray and neutron diffraction (black diamonds)~\cite{hunt01}.
  			The onset temperature for spin-stripe order, $T_{\rm spin}$, is detected 
  			by neutron diffraction (black down triangles;~\cite{ichikawa00}). 
  			We can then see that $T_{\rm \rho} \approx 2 T_{\rm charge}$. The superconducting 
  			transition temperature $T_{\rm c}$ is shown as open black circles \cite{daou08,ichikawa00}. 
  			All lines are a guide to the eye.}
		\label{fig1}
	\end{center}
\end{figure}

In this Letter, we present measurements of the $c$-axis resistivity $\rho_{\rm c}$ 
in Nd-LSCO as a function of temperature $T$ in the vicinity of the QCP. 
These out-of-plane measurements reveal the same behavior as found in the in-plane data, 
namely a linear-$T$ resistivity down to a temperature $T_{\rm \rho}$ below which $\rho(T)$ 
starts to deviate upwards~\cite{daou08}. In-plane data at $p=0.20$ and $p=0.24$ show that 
$T_{\rm \rho}$ goes to zero between these two dopings (see Fig.~\ref{fig1}). 
Here we use $c$-axis samples at intermediate dopings to pin down with greater accuracy 
the critical doping $p^\star$ where $T_{\rm \rho} \to 0$. 

The four samples of Nd-LSCO used in this study were grown at the University of Texas, 
as described elsewhere~\cite{daou08}.
They have a doping of $p = 0.20, 0.22, 0.23$ and $0.24$, respectively. 
The resistivity $\rho_{\rm c}$ was measured at the National High Magnetic Field Laboratory
in Tallahassee, in steady magnetic fields up to $45$\,T.

\begin{figure}
	\begin{center}	
		\epsfig{figure=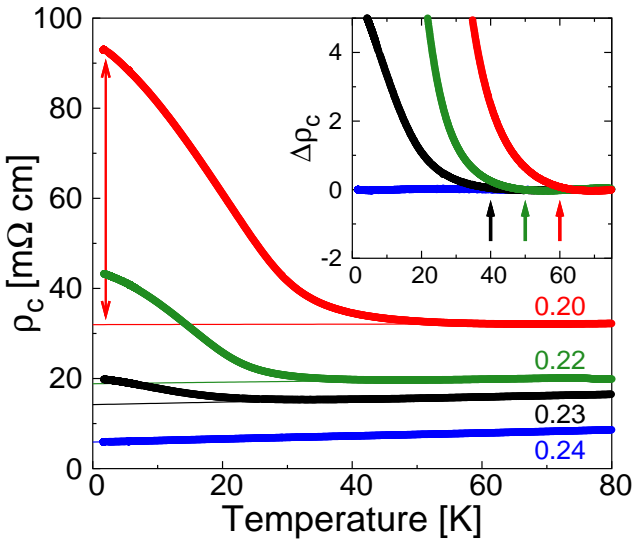, angle=270,width=0.46\textwidth}
			\caption{(Color online) $c$-axis resistivity $\rho_{\rm c}$ of Nd-LSCO at 
				$p = 0.20, 0.22, 0.23$ and $0.24$ (top to bottom) as a function of 
				temperature $T$, in a magnetic field of $45$\,T (applied along the $c$-axis). 
				The lines are a linear fit of the data between 60 and 80 K for each curve separately.
				We define as $\Delta\rho_{\rm c}(T)$ the difference between data and fit, plotted in the inset.
				In the limit $T \to 0$, this gives us $\Delta\rho_{\rm c}(0)$, shown by the double-headed (red) arrow.
				This quantity is plotted in Fig.~\ref{fig3}. 
				Inset: $\Delta\rho_{\rm c}$ vs $T$ for the four dopings. The onset of the upturn at 
				a temperature $T_{\rm \rho}$ is detected as an upward deviation from zero. 
				Arrows mark $T_{\rm \rho} = 60 \pm 10, 50 \pm 10$ and $40 \pm 10$\,K for 
				$p = 0.20$, $0.22$ and $0.23$ respectively. At $p = 0.24$, $T_{\rm \rho}$ = 0.}
		\label{fig2}
	\end{center}
\end{figure}
\begin{figure}
	\begin{center}	
		\epsfig{figure=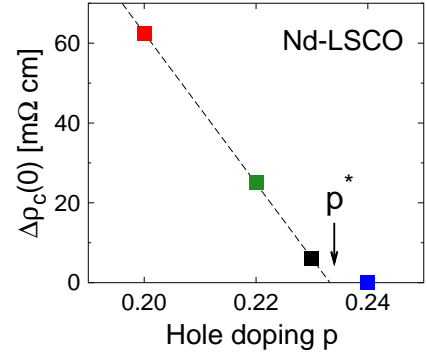, angle=270,width=0.3\textwidth}
			\caption{(Color online) Height of the upturn in $\rho_{\rm c}(T)$ 
			as a function of doping $p$, defined as the $T \to 0$ limit of 
			$\Delta\rho_{\rm c}(T)$, plotted in Fig.~\ref{fig2}.
 			A linear extrapolation (dashed line) of the $\Delta\rho_{\rm c}(0)$ 
 			data at $p = 0.20$, 0.22 and 0.23 yields $p^\star = 0.235 \pm 0.005$ 
 			as the critical doping beyond which $\rho_{\rm c}(T)$ show no upturn. 
 			This is the QCP below which Fermi-surface reconstruction occurs.}
		\label{fig3}
	\end{center}
\end{figure}

Our $\rho_{\rm c}$ data in a field of $45$\,T is presented in Fig.~\ref{fig2}. 
(Note that there is negligible magneto-resistance in all cases.)
A linear fit to the data below 80 K is shown as a solid line. In the inset, 
we plot $\Delta\rho_{\rm c}$, the difference between data and fit. At $p=0.24$, 
we see that $\rho_{\rm c}(T)$ remains linear down to the lowest temperature, 
as previously reported~\cite{daou08}. At lower doping, an upturn is observed 
below a temperature $T_{\rm \rho}$ (see arrows in inset) which is plotted vs doping 
in Fig.~\ref{fig1} (red circles). The values of $T_{\rm \rho}$ obtained from $\rho_{\rm c}$ 
are seen to agree well with the overall doping dependence of $T_{\rm \rho}$ obtained from 
$\rho_{\rm  ab}$ (reproduced from ref.~\cite{daou08}).

Another way to describe the evolution of the $c$-axis resistivity data is to plot 
the magnitude of the upturn as a function of doping, defined as $\Delta\rho_{\rm c}(0)$, 
the difference between data and fit in the limit of  $T \to 0$ (red double-headed arrow in Fig.~\ref{fig2}).
Fig.~\ref{fig3} shows $\Delta\rho_{\rm c}(0)$ as a function of doping. It is clear that 
the height of the upturn goes down as the doping is increased, extrapolating to zero at $p = 0.235 \pm 0.005$.
This accurately locates the quantum critical point below which Fermi-surface reconstruction begins. 
We infer that this is where translational symmetry is broken at $T = 0$.

As noted previously from in-plane data~\cite{daou08,taillefer09}, the upturn begins 
at a temperature significantly above the ordering temperature for stripe order, 
at $T_{\rm charge}$ (see Fig.~\ref{fig1}), with $T_\rho \approx 2 T_{\rm charge}$. 
This suggests a two-step transformation of the electronic behaviour~\cite{taillefer09}: a first transformation 
at high temperature, detected in the resistivity and the quasiparticle Nernst signal below 
$T_\nu$~\cite{cyrchoiniere09}, with $T_{\rm \rho} \simeq T_\nu$, and a second transformation at 
the stripe ordering temperature, detected in the Hall~\cite{daou08} and Seebeck coefficients~\cite{daou09}. 

We propose that the temperature $T_{\rm \rho} \simeq T_\nu$ is in fact the pseudogap temperature $T^\star$.  
The pseudogap phase would then most likely be a fluctuating precursor of the spin/charge density wave 
(stripe) order observed at lower temperature (below $T_{\rm charge}$). This QCP may be a generic feature 
of hole-doped cuprates. Indeed, recent measurements of $c$-axis resistivity in overdoped Bi-2212 crystals 
show upturns below a temperature $T_\rho$ which also goes to zero at $p \simeq 0.24$~\cite{murata}. 
Moreover, a reconstruction of the Fermi surface by stripe-like order may also be a more general occurrence, 
given the very similar anomalies observed in YBCO in both the Hall~\cite{taillefer09} and Seebeck~\cite{chang09} coefficients.

\end{document}